\documentclass[10pt,final,journal,twocolumn]{IEEEtran}
\ifCLASSINFOpdf

\else
\usepackage{algorithm}
\usepackage{booktabs}

\fi
\usepackage{cite}
\usepackage[cmex10]{amsmath}
\usepackage{amsthm}
\usepackage{stfloats}
\usepackage{multicol,multienum}
\usepackage{multirow}

\usepackage{amssymb}
\usepackage{url}
\usepackage{amsmath}
\usepackage{xcolor}
\usepackage[dvips]{graphicx}
\usepackage{subfigure}
\usepackage{caption}
\usepackage{amsmath}
\usepackage{amsfonts,amssymb}
\usepackage{bbm}
\usepackage[T1]{fontenc}
\usepackage{algorithm}
\usepackage{algpseudocode}
\usepackage{ulem}

\usepackage[linesnumbered, ruled]{}
\makeatletter
\renewcommand{\maketag@@@}[1]{\hbox{\m@th\normalsize\normalfont#1}}%

\makeatother

\begin{document}
\hyphenation{op-tical net-works semi-conduc-tor}
\title{\LARGE \vspace{-0.4em} Movable Antennas-Enabled Two-User Multicasting: Do We Really Need Alternating Optimization for Minimum Rate Maximization?}
\author{Guojie Hu, Qingqing Wu,~\textit{Senior Member}, \textit{IEEE}, Donghui Xu, Kui Xu,~\textit{Member}, \textit{IEEE}, Jiangbo Si,~\textit{Senior Member}, \textit{IEEE}, Yunlong Cai,~\textit{Senior Member}, \textit{IEEE}, and Naofal Al-Dhahir,~\textit{Fellow}, \textit{IEEE}\vspace{-2.1em}
\thanks{
%
Guojie Hu and Donghui Xu are with the College of Communication Engineering, Rocket Force University of Engineering, Xi'an 710025, China (lgdxhgj@sina.com). Qingqing Wu is with the Department of Electronic Engineering, Shanghai Jiao Tong University, Shanghai 200240, China. Kui Xu is with the College of Communications Engineering, Army Engineering University of PLA, Nanjing 210007, China. Jiangbo Si is with the Integrated Service Networks Lab of Xidian University, Xi'an 710100, China. Yunlong Cai is with the College of Information Science and Electronic Engineering, Zhejiang University, Hangzhou 310027, China. Naofal Al-Dhahir is with the Department of Electrical and Computer Engineering, The University of Texas at Dallas, Richardson, TX 75080 USA.
}
}
\IEEEpeerreviewmaketitle
\maketitle
\begin{abstract}
Movable antenna (MA) technology, which can reconfigure wireless channels by flexibly moving antenna positions in a specified region, has great potential for improving communication performance. In this paper, we consider a new setup of MAs-enabled multicasting, where we adopt a simple setting in which a linear MA array-enabled source (${\rm{S}}$) transmits a common message to two single-antenna users ${\rm{U}}_1$ and ${\rm{U}}_2$. We aim to maximize the minimum rate among these two users, by jointly optimizing the transmit beamforming and antenna positions at ${\rm{S}}$. Instead of utilizing the widely-used alternating optimization (AO) approach, we reveal, with rigorous proof, that the above two variables can be optimized separately: i) the optimal antenna positions can be firstly determined via the successive convex approximation technique, based on the rule of maximizing the correlation between ${\rm{S}}$-${\rm{U}}_1$ and ${\rm{S}}$-${\rm{U}}_2$ channels; ii) afterwards, the optimal closed-form transmit beamforming can be derived via simple arguments. Compared to AO, this new approach yields the same performance but reduces the computational complexities significantly. Moreover, it can provide insightful conclusions which are not possible with AO.
\end{abstract}

\IEEEpeerreviewmaketitle
\vspace{-15pt}
\section{Introduction}
\vspace{-5pt}
Beamforming, originating from the multiple-antenna technique, plays an important role in the area of improving signal receiving qualities or eliminating unfavorable interferences in 3G$-$B5G systems \cite{Xiaozhenyu}.

Current beamforming is implemented by leveraging fixed-position antennas (FPAs), where antennas cannot adjust their positions. This setting, undeniably, brings a fundamental limitation. Specifically, the design of beamforming is significantly related to wireless channels, which, however, cannot be reconfigured with FPAs. Then, if channel conditions are not desirable, beamforming with FPAs may not be able to attain its full potential \cite{SPL}. Motivated by this observation, one natural question arises, namely., whether wireless channels can be reconfigured to cater to effective beamforming designs? Recently, a promising technique called movable antennas (MAs) may provide the answer \cite{Zhulipeng_CM, FA1}. Using MAs, antenna positions at transmitters/receivers can be adjusted in a local region using the stepper motors or servos. This flexible behavior of MAs reshapes wireless channels adaptively, and thus brings an additional spatial degree of freedom (DoF) for further improving the system performance.

 \begin{figure}
\centering
\includegraphics[width=6.5cm]{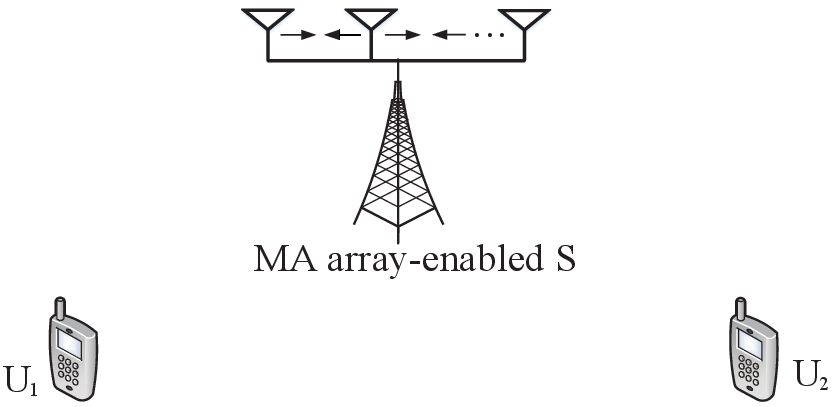}
\captionsetup{font=small}
\caption{Illustration of the system model.} \label{fig:Fig1}
\vspace{-22pt}
\end{figure}

Inspired by its potential merits, early works have integrated MAs into numerous setups such as multi-input multi-out system \cite{Mawenyan_MIMO,Gao_Xiqi_CL,Yongpeng_WU_Glob}, multi-user uplink/downlink communications \cite{Pixiangyu_GC,Guojie_Fluid,Songjie_WCL,Yifei_Wu_arxiv,gao2024joint,10416896}, interference networks \cite{Wanghonghao_arxiv1,Zhulipeng_CL}, physical-layer security systems \cite{Guojie_SPL,Zhenqiao,Guojie_TMC}, computation networks \cite{MA_computation}, coordinate multi-point (CoMP) reception systems \cite{Guojie_COMP} and so on. Different from previous works, in this paper we consider a new setup of MAs-enabled two-user multicasting, where a linear MA array-enabled source (${\rm{S}}$) transmits a common message to two single-antenna users ${\rm{U}}_1$ and ${\rm{U}}_2$. The objective is to maximize the minimum rate among these two users, by jointly optimizing the transmit beamforming and antenna positions at ${\rm{S}}$. Generally, the above two variables are coupled with each other, making the problem highly non-convex. Therefore, like most related works, alternating optimization (AO), which alternately handles beamforming (antenna positions) given antenna positions (beamforming) being fixed in each iteration, is the most conventional solution. However, AO requires high computational complexities because it involves numerous outer iterations, which may be undesirable for practical implementations. Facing this drawback, we try to answer one fundamental problem, i.e., whether the above two variables can be optimized separately? Motivated by the key logic that \textit{the optimization for antenna positions is actually creating favorable prerequisites for beamforming}, we reveal, with rigorous proof, that antenna positions can be first optimized based on the rule of maximizing the correlation between ${\rm{S}}$-${\rm{U}}_1$ and ${\rm{S}}$-${\rm{U}}_2$ channels, the process of which has no relation with beamforming; afterwards, the optimal transmit beamforming can be obtained. Since no outer iterations are involved, our proposed scheme enjoys lower complexities but achieves the same performance as compared to AO. More importantly, the idea of adjusting antenna positions from the perspective of varying channel correlation(s) may establish a new optimization flow in most MAs-enabled systems.

{\textit{{Notations:}} For a complex number $a$, ${\mathop{\rm Re}\nolimits} \left[ a \right]$ is its real part. For a complex vector ${\bf{h}}$, ${{\bf{h}}^T}$, ${{\bf{h}}^*}$, ${{\bf{h}}^H}$ and $\left\| {\bf{h}} \right\|$ denote its transpose, conjugate, conjugate transpose and Frobenius norm, respectively. ${\bf{I}}$ is the identity matrix, ${\prod _{\bf{X}}} = {\bf{X}}{({{\bf{X}}^H}{\bf{X}})^{ - 1}}{{\bf{X}}^H}$ is the orthogonal projection onto the column space of ${\bf{X}}$, with ${( \cdot )^{ - 1}}$ denoting the inverse operation, and $\prod _{\bf{X}}^ \bot  = {\bf{I}} - {\prod _{\bf{X}}}$ is the orthogonal projection onto the orthogonal complement of the column space of ${\bf{X}}$.

 %


 \newcounter{mytempeqncnt}
 \vspace{-5pt}
\section{System Model and Problem Formulation}
As shown in Fig. 1, we consider the MA-enabled two-user downlink multicasting, where the source $\rm{S}$ aims to transmit a common message to two users ${\rm{U}}_1$ and ${\rm{U}}_2$. $\rm{S}$ is equipped with a linear MA array consisting of $N \ge 2$ antennas, while ${\rm{U}}_1$ and ${\rm{U}}_2$ are equipped with a single antenna each.

As the linear array is employed at S, the one-dimensional positions of $N$ MAs relative to the reference point zero are denoted as ${\left[ {{x_1},{x_2},...,{x_N}} \right]^T} \buildrel \Delta \over = {\bf{x}}$. Without loss of generality, let $0 \le {x_1} < {x_2} < ... < {x_N} \le L$, where $L$ is the total span for the movement region of MAs. To simplify analysis and without affecting the obtained conclusions, we consider the line-of-sight channel model for $\rm{S}$-${\rm{U}}_1$ and $\rm{S}$-${\rm{U}}_2$ links \cite{Zhulipeng_CL}. Then, given ${\bf{x}}$, the channel vector from $\rm{S}$ to ${\rm{U}}_i$, $i = 1, 2$, can be expressed as
\setlength\abovedisplayskip{1.2pt}
\setlength\belowdisplayskip{1.2pt}
\begin{equation}
\begin{split}{}
{{\bf{h}}_{s{u_i}}}({\bf{x}}) = \frac{1}{{\sqrt {d_{s{u_i}}^\tau } }}{\left[ {{e^{j\frac{{2\pi }}{\lambda }{x_1}\sin {\theta _{s{u_i}}}}},...,{e^{j\frac{{2\pi }}{\lambda }{x_N}\sin {\theta _{s{u_i}}}}}} \right]^T},
\end{split}
\end{equation}
where ${d_{s{u_i}}}$ denotes the distance between $\rm{S}$ and ${\rm{U}}_i$, $\tau $ is the path loss exponent, ${{\theta _{s{u_i}}}}$ is the angle of departure (AoD) from the linear array at $\rm{S}$ to ${\rm{U}}_i$ and $\lambda $ is the carrier wavelength. 

Given ${\bf{x}}$ and further denoting the digital beamforming vector of $\rm{S}$ as ${\bf{w}} \in {{\mathbb{C}}^{N \times 1}}$, with $\left\| {\bf{w}} \right\|_2^2 = 1$, the received signal-to-noise ratio (SNR) at ${\rm{U}}_i$ can be derived as
\begin{equation}
\begin{split}{}
{\gamma _{{u_i}}}({\bf{w}},{\bf{x}}) = \frac{{{P_s}}}{{{\sigma ^2}}}{\left| {{\bf{h}}_{s{u_i}}^T({\bf{x}}){\bf{w}}} \right|^2} = \frac{{{P_s}}}{{d_{s{u_i}}^\tau {\sigma ^2}}}{\left| {\overline {\bf{h}} _{s{u_i}}^T({\bf{x}}){\bf{w}}} \right|^2},
\end{split}
\end{equation}
where $P_s$ is the transmit power of S, ${\sigma ^2}$ is the noise power and ${\overline {\bf{h}} _{s{u_i}}}({\bf{x}}) = {\left[ {{e^{j\frac{{2\pi }}{\lambda }{x_1}\sin {\theta _{s{u_i}}}}},...,{e^{j\frac{{2\pi }}{\lambda }{x_N}\sin {\theta _{s{u_i}}}}}} \right]^T}$.

In this paper, considering the fairness issue, we aim to maximize the minimum SNR $\min \left( {{\gamma _{{u_1}}}({\bf{w}},{\bf{x}}),{\gamma _{{u_2}}}({\bf{w}},{\bf{x}})} \right)$ (or equivalently, the minimum rate ${\log _2}\left( {1 + \min \left( {{\gamma _{{u_1}}}({\bf{w}},{\bf{x}}),{\gamma _{{u_2}}}({\bf{w}},{\bf{x}})} \right)} \right)$) among these two users, by jointly optimizing the transmit beamforming ${\bf{w}}$ and antenna positions ${\bf{x}}$ at $\rm{S}$. Hence, the optimization problem can be formulated as
 \begin{align}
&({\rm{P1}}):{\rm{  }}\mathop {\max }\limits_{{\bf{w}}, {{\bf{x}}}} \ \min \left( {{\gamma _{{u_1}}}({\bf{w}},{\bf{x}}),{\gamma _{{u_2}}}({\bf{w}},{\bf{x}})} \right)\tag{${\rm{3a}}$}\\
{\rm{              }}&\ {\rm{s.t.}} \quad \left\| {\bf{w}} \right\|_2^2 = 1,\tag{${\rm{3b}}$}\\
 &\quad \quad \quad {x_n} - {x_{n - 1}} \ge {d_{\min }},n = 2,3,...,N,\tag{${\rm{3c}}$}\\
&\quad \quad \ \ \left\{ {{x_n}} \right\}_{n = 1}^N \in [0,L].\tag{${\rm{3d}}$}
 \end{align}
where $d_{{\rm{min}}}$ in (3c) is the minimum distance between any two adjacent MAs to avoid the coupling effect.

Observing the structure of the objective in (3a), clearly ${\bf{w}}$ and ${\bf{x}}$ are coupled with each other in the terms $\overline {\bf{h}} _{s{u_1}}^T({\bf{x}}){\bf{w}}$ and $\overline {\bf{h}} _{s{u_2}}^T({\bf{x}}){\bf{w}}$, making (P1) highly non-convex. Generally, to solve (P1), AO, which alternately optimizes ${\bf{w}}$ and ${\bf{x}}$ given the other variable being fixed, is a widely adopted approach in the literature \cite{Mawenyan_MIMO, Guojie_SPL,MA_computation}. Nevertheless, since numerous iterations are involved in AO, this method is generally computationally inefficient. Therefore, one fundamental question has arisen, i.e., is it possible that these two variables can be optimized separately? We will answer this question in the next section.
\vspace{-7pt}
\section{Solving (P1) without AO}
To proceed, given ${\bf{x}}$, since the design for ${\bf{w}}$ needs to balance the values of ${\left| {\overline {\bf{h}} _{s{u_1}}^T({\bf{x}}){\bf{w}}} \right|^2}$ and ${\left| {\overline {\bf{h}} _{s{u_2}}^T({\bf{x}}){\bf{w}}} \right|^2}$ for achieving the optimal trade-off, it is well-known that the optimal ${\bf{w}}$ enjoys the following structure \cite{Han_Liang}
\begin{equation}
\setcounter{equation}{4}
{\bf{w}} = t\frac{{{\prod _{\overline {\bf{h}} _{s{u_1}}^*({\bf{x}})}}\overline {\bf{h}} _{s{u_2}}^*({\bf{x}})}}{{\left\| {{\prod _{{{\overline {\bf{h}} }_{s{u_1}}}({\bf{x}})}}{{\overline {\bf{h}} }_{s{u_2}}}({\bf{x}})} \right\|}} + \sqrt {1 - {t^2}} \frac{{\prod _{\overline {\bf{h}} _{s{u_1}}^*({\bf{x}})}^ \bot \overline {\bf{h}} _{s{u_2}}^*({\bf{x}})}}{{\left\| {\prod _{{{\overline {\bf{h}} }_{s{u_1}}}({\bf{x}})}^ \bot {{\overline {\bf{h}} }_{s{u_2}}}({\bf{x}})} \right\|}},
\end{equation}
where $t$ is a real number that can be adjusted in the range of $[0,1]$. Substituting (4) into (2), $\min \left( {{\gamma _{{u_1}}}({\bf{w}},{\bf{x}}),{\gamma _{{u_2}}}({\bf{w}},{\bf{x}})} \right)$ can be simplified into a function related to $t$ and ${\bf{x}}$, i.e.,
\begin{equation}
\begin{split}{}
&\min \left( {{\gamma _{{u_1}}}(t,{\bf{x}}),{\gamma _{{u_2}}}(t,{\bf{x}})} \right) \\
=&\min \left( {\begin{array}{*{20}{l}}
{\frac{{{P_s}}}{{d_{s{u_1}}^\tau {\sigma ^2}}}{a^2}({\bf{x}}){t^2},}\\
{\frac{{{P_s}}}{{d_{s{u_2}}^\tau {\sigma ^2}}}{{\left( {b({\bf{x}})t + c({\bf{x}})\sqrt {1 - {t^2}} } \right)}^2}}
\end{array}} \right),
\end{split}
\end{equation}
where based on \cite{Han_Liang}, we have
\begin{equation} \nonumber
\begin{split}{}
a({\bf{x}}) =& \frac{{\left| {\overline {\bf{h}} _{s{u_1}}^T({\bf{x}}){\prod _{\overline {\bf{h}} _{s{u_1}}^*({\bf{x}})}}\overline {\bf{h}} _{s{u_2}}^*({\bf{x}})} \right|}}{{\left\| {{\prod _{{{\overline {\bf{h}} }_{s{u_1}}}({\bf{x}})}}{{\overline {\bf{h}} }_{s{u_2}}}({\bf{x}})} \right\|}}, \\
b({\bf{x}}) =& \left\| {\prod\nolimits_{{{\overline {\bf{h}} }_{s{u_1}}}({\bf{x}})} {{{\overline {\bf{h}} }_{s{u_2}}}({\bf{x}})} } \right\|, c({\bf{x}}) = \left\| {\prod\nolimits_{{{\overline {\bf{h}} }_{s{u_1}}}({\bf{x}})}^ \bot  {{{\overline {\bf{h}} }_{s{u_2}}}({\bf{x}})} } \right\|.
\end{split}
\end{equation}

 \begin{figure*}[b!]
 \vspace{-8pt}
  \hrulefill
\setcounter{mytempeqncnt}{\value{equation}}
\begin{equation} \nonumber
\begin{split}{}
a({\bf{x}}) = \frac{{\left| {\overline {\bf{h}} _{s{u_1}}^T({\bf{x}})\overline {\bf{h}} _{s{u_1}}^*({\bf{x}}){{\left( {{{\left( {\overline {\bf{h}} _{s{u_1}}^*({\bf{x}})} \right)}^H}\overline {\bf{h}} _{s{u_1}}^*({\bf{x}})} \right)}^{ - 1}}{{\left( {\overline {\bf{h}} _{s{u_1}}^*({\bf{x}})} \right)}^H}\overline {\bf{h}} _{s{u_2}}^*({\bf{x}})} \right|}}{{\left\| {{{\overline {\bf{h}} }_{s{u_1}}}({\bf{x}}){{\left( {\overline {\bf{h}} _{s{u_1}}^H({\bf{x}}){{\overline {\bf{h}} }_{s{u_1}}}({\bf{x}})} \right)}^{ - 1}}\overline {\bf{h}} _{s{u_1}}^H({\bf{x}}){{\overline {\bf{h}} }_{s{u_2}}}({\bf{x}})} \right\|}}\mathop {{\rm{  }} = }\limits^{(1)} \frac{{\left| {\overline {\bf{h}} _{s{u_1}}^T({\bf{x}})\overline {\bf{h}} _{s{u_1}}^*({\bf{x}}){{\left( {\overline {\bf{h}} _{s{u_1}}^*({\bf{x}})} \right)}^H}\overline {\bf{h}} _{s{u_2}}^*({\bf{x}})} \right|/N}}{{\left| {\overline {\bf{h}} _{s{u_1}}^H({\bf{x}}){{\overline {\bf{h}} }_{s{u_2}}}({\bf{x}})} \right|\left\| {{{\overline {\bf{h}} }_{s{u_1}}}({\bf{x}})} \right\|/N}}\mathop {{\rm{  }} = }\limits^{(2)} \sqrt N .
\end{split}
\end{equation}
\setcounter{equation}{\value{mytempeqncnt}}
\vspace{-12pt}
\end{figure*}
  \begin{figure*}[b!]
\setcounter{mytempeqncnt}{\value{equation}}
\begin{equation} \nonumber
\begin{split}{}
b({\bf{x}}) = \left\| {{{\overline {\bf{h}} }_{s{u_1}}}({\bf{x}}){{\left( {\overline {\bf{h}} _{s{u_1}}^H({\bf{x}}){{\overline {\bf{h}} }_{s{u_1}}}({\bf{x}})} \right)}^{ - 1}}\overline {\bf{h}} _{s{u_1}}^H({\bf{x}}){{\overline {\bf{h}} }_{s{u_2}}}({\bf{x}})} \right\| = \left| {\overline {\bf{h}} _{s{u_1}}^H({\bf{x}}){{\overline {\bf{h}} }_{s{u_2}}}({\bf{x}})} \right|\left\| {{{\overline {\bf{h}} }_{s{u_1}}}({\bf{x}})} \right\|/N = \left| {\overline {\bf{h}} _{s{u_1}}^H({\bf{x}}){{\overline {\bf{h}} }_{s{u_2}}}({\bf{x}})} \right|/\sqrt N .
\end{split}
\end{equation}
\setcounter{equation}{\value{mytempeqncnt}}
\vspace{-12pt}
\end{figure*}
  \begin{figure*}[b!]
\setcounter{mytempeqncnt}{\value{equation}}
\begin{equation} \nonumber
\begin{split}{}
c({\bf{x}}) \mathop  = \limits^{(3)} & \sqrt {\left\| {{{\overline {\bf{h}} }_{s{u_2}}}({\bf{x}})} \right\|_2^2 + \left\| {{{\overline {\bf{h}} }_{s{u_1}}}({\bf{x}}){{\left( {\overline {\bf{h}} _{s{u_1}}^H({\bf{x}}){{\overline {\bf{h}} }_{s{u_1}}}({\bf{x}})} \right)}^{ - 1}}\overline {\bf{h}} _{s{u_1}}^H({\bf{x}}){{\overline {\bf{h}} }_{s{u_2}}}({\bf{x}})} \right\|_2^2 - 2{\rm{Re}}\left[ {\overline {\bf{h}} _{s{u_2}}^H({\bf{x}}){{\overline {\bf{h}} }_{s{u_1}}}({\bf{x}})\overline {\bf{h}} _{s{u_1}}^H({\bf{x}}){{\overline {\bf{h}} }_{s{u_2}}}({\bf{x}})} \right]/N}  \nonumber \\
 =& \sqrt {N + {{\left| {\overline {\bf{h}} _{s{u_1}}^H({\bf{x}}){{\overline {\bf{h}} }_{s{u_2}}}({\bf{x}})} \right|}^2}/N - 2{{\left| {\overline {\bf{h}} _{s{u_2}}^H({\bf{x}}){{\overline {\bf{h}} }_{s{u_1}}}({\bf{x}})} \right|}^2}/N}  = \sqrt {N - {{\left| {\overline {\bf{h}} _{s{u_1}}^H({\bf{x}}){{\overline {\bf{h}} }_{s{u_2}}}({\bf{x}})} \right|}^2}/N}.
\end{split}
\end{equation}
\setcounter{equation}{\value{mytempeqncnt}}
\vspace{-12pt}
\end{figure*}
  \begin{figure*}[b!]
  \hrulefill
\setcounter{mytempeqncnt}{\value{equation}}
\setcounter{equation}{5}
\begin{equation}
\begin{split}{}
\min \left( {{\gamma _{{u_1}}}(t,{\bf{x}}),{\gamma _{{u_2}}}(t,{\bf{x}})} \right) = \min \left( {\frac{{{P_s}N}}{{d_{s{u_1}}^\tau {\sigma ^2}}}{t^2},\frac{{{P_s}}}{{d_{s{u_2}}^\tau {\sigma ^2}}}{{\left( {\frac{{f({\bf{x}})}}{{\sqrt N }}t + \sqrt {N - \frac{{{f^2}({\bf{x}})}}{N}} \sqrt {1 - {t^2}} } \right)}^2}} \right) \buildrel \Delta \over = \Theta (t,{\bf{x}}).
\end{split}
\end{equation}
\setcounter{equation}{\value{mytempeqncnt}}
\vspace{-12pt}
\end{figure*}

Since the current expressions of $a({\bf{x}})$, $b({\bf{x}})$ and $c({\bf{x}})$ in \cite{Han_Liang} are overly complex, we are motivated to further simplify them at the bottom of next page, where the equality (1) is established since ${\left( {{{\left( {\overline {\bf{h}} _{s{u_1}}^*({\bf{x}})} \right)}^H}\overline {\bf{h}} _{s{u_1}}^*({\bf{x}})} \right)^{ - 1}} = 1/N$, the equality (2) is established since $\left\| {{{\overline {\bf{h}} }_{s{u_1}}}({\bf{x}})} \right\| = \sqrt N $, and the equality (3) is established since $\left\| {{\bf{u}} - {\bf{v}}} \right\| = \sqrt {\left\| {\bf{u}} \right\|_2^2 + \left\| {\bf{v}} \right\|_2^2 - 2{\mathop{\rm Re}\nolimits} \left[ {{{\bf{u}}^H}{\bf{v}}} \right]} $ for any ${\bf{u}},{\bf{v}} \in {{\mathbb{C}}^{N \times 1}}$. Now, substituting the simplified $a({\bf{x}})$, $b({\bf{x}})$ and $c({\bf{x}})$ into (5), $\min \left( {{\gamma _{{u_1}}}(t,{\bf{x}}),{\gamma _{{u_2}}}(t,{\bf{x}})} \right) \buildrel \Delta \over = \Theta (t,{\bf{x}})$ can be simplified as shown in (6) at the bottom of next page, where $f({\bf{x}}) = \left| {\overline {\bf{h}} _{s{u_1}}^H({\bf{x}}){{\overline {\bf{h}} }_{s{u_2}}}({\bf{x}})} \right|$.

\textbf{Proposition 1:} Given ${\bf{x}}$, i.e, $f({\bf{x}})$, to maximize $\Theta (t,{\bf{x}})$, the optimal $t$ must lie in the range of $\left[ {\frac{{f({\bf{x}})}}{N},1} \right]$.
\begin{proof}
Focusing on $\Theta (t,{\bf{x}})$, given ${f({\bf{x}})}$, ${\frac{{{P_s}N}}{{d_{{s{u_1}}}^\tau {\sigma ^2}}}{t^2}}$ is monotonically increasing with respect to (w.r.t.) $t \in \left[ {0,1} \right]$, while using the derivative method, it is determined that ${\frac{{{P_s}}}{{d_{s{u_2}}^\tau {\sigma ^2}}}{{\left( {\frac{{f({\bf{x}})}}{{\sqrt N }}t + \sqrt {N - \frac{{{f^2}({\bf{x}})}}{N}} \sqrt {1 - {t^2}} } \right)}^2}}$ is monotonically increasing when $t \in \left[ {0,\frac{{f({\bf{x}})}}{N}} \right]$ and monotonically decreasing when $t \in \left[ {\frac{{f({\bf{x}})}}{N},1} \right]$. Hence, only when $t \in \left[ {\frac{{f({\bf{x}})}}{N},1} \right]$, the optimal trade-off between ${\frac{{{P_s}N}}{{d_{{s{u_1}}}^\tau {\sigma ^2}}}{t^2}}$ and ${\frac{{{P_s}}}{{d_{s{u_2}}^\tau {\sigma ^2}}}{{\left( {\frac{{f({\bf{x}})}}{{\sqrt N }}t + \sqrt {N - \frac{{{f^2}({\bf{x}})}}{N}} \sqrt {1 - {t^2}} } \right)}^2}}$ can be achieved. This completes the proof.
\end{proof}

Based on the above analysis, (P1) can be equivalently reformulated as
 \begin{align}
&({\rm{P1.1}}):{\rm{  }}\mathop {\max }\limits_{t, {{\bf{x}}}} \ \Theta (t,{\bf{x}})\tag{${\rm{7a}}$}\\
{\rm{              }}&\ {\rm{s.t.}} \quad \quad t \in \left[ {f({\bf{x}})/N,1} \right],\tag{${\rm{7b}}$}\\
 &\quad \quad \quad \ (3{\rm{c}}), (3{\rm{d}}).\tag{${\rm{7c}}$}
 \end{align}

Unfortunately, the variables $t$ and ${\bf{x}}$ are also coupled with each other in $\Theta (t,{\bf{x}})$, making the simplified problem (P1.1) still non-convex. Nevertheless, $t$ and ${\bf{x}}$ can be optimized separately, by leveraging the following important Proposition.

\textbf{Proposition 2:} The minimum rate reaches its maximum if and only if ${f({\bf{x}})}$ achieves its maximum.
\begin{proof}
Considere the expression of $\Theta (t,{\bf{x}})$, and let $g(t,{\bf{x}}) = \frac{{f({\bf{x}})}}{{\sqrt N }}t + \sqrt {N - \frac{{{f^2}({\bf{x}})}}{N}} \sqrt {1 - {t^2}} $. Denote two patterns of antenna positions as ${{\bf{x}}_1}$ and ${{\bf{x}}_2}$, and let $f({{\bf{x}}_1}) \ge f({{\bf{x}}_2})$. Then, for these two patterns, the corresponding $t$ should lie in $\left[ {\frac{{f({{\bf{x}}_1})}}{{\sqrt N }},1} \right]$ and $\left[ {\frac{{f({{\bf{x}}_2})}}{{\sqrt N }},1} \right]$, respectively. Further note that $g(t,{{\bf{x}}_1})$ and $g(t,{{\bf{x}}_2})$ are monotonically decreasing w.r.t. $t \in \left[ {\frac{{f({{\bf{x}}_1})}}{{\sqrt N }},1} \right]$ and $t \in \left[ {\frac{{f({{\bf{x}}_2})}}{{\sqrt N }},1} \right]$, respectively, and $g\left( {\frac{{f({{\bf{x}}_1})}}{{\sqrt N }},{{\bf{x}}_1}} \right) = g\left( {\frac{{f({{\bf{x}}_2})}}{{\sqrt N }},{{\bf{x}}_2}} \right) = \sqrt N $. Then: i) for any $t \in \left[ {\frac{{f({{\bf{x}}_2})}}{{\sqrt N }},\frac{{f({{\bf{x}}_1})}}{{\sqrt N }}} \right]$, $g\left( {t,{{\bf{x}}_2}} \right) \le g\left( {\frac{{f({{\bf{x}}_1})}}{{\sqrt N }},{{\bf{x}}_1}} \right)$, leading to $\Theta (t,{{\bf{x}}_2}) \le \Theta \left( {\frac{{f({{\bf{x}}_1})}}{{\sqrt N }},{{\bf{x}}_1}} \right)$; ii) for any $t \in \left[ {\frac{{f({{\bf{x}}_1})}}{N},1} \right]$, we can derive that
\setlength\abovedisplayskip{1.2pt}
\setlength\belowdisplayskip{1.2pt}
\begin{equation} \nonumber
\begin{split}{}
&\quad g\left( {t,{{\bf{x}}_1}} \right) - g\left( {t,{{\bf{x}}_2}} \right)\\
 &= \left( {\frac{{f({{\bf{x}}_1})}}{{\sqrt N }} - \frac{{f({{\bf{x}}_2})}}{{\sqrt N }}} \right)t\\
 &+ \left( {\sqrt {N - \frac{{{f^2}({{\bf{x}}_1})}}{N}}  - \sqrt {N - \frac{{{f^2}({{\bf{x}}_2})}}{N}} } \right)\sqrt {1 - {t^2}}
\end{split}
\end{equation}
\begin{equation} \nonumber
\begin{split}{}
&\mathop  \ge \limits^{(4)} \left( {\frac{{f({{\bf{x}}_1})}}{{\sqrt N }} - \frac{{f({{\bf{x}}_2})}}{{\sqrt N }}} \right)\frac{{f({{\bf{x}}_1})}}{N}\\
 &+ \left( {\sqrt {N - \frac{{{f^2}({{\bf{x}}_1})}}{N}}  - \sqrt {N - \frac{{{f^2}({{\bf{x}}_2})}}{N}} } \right)\sqrt {1 - \frac{{{f^2}({{\bf{x}}_1})}}{{{N^2}}}} \\
 &= \frac{1}{{\sqrt N }}\left( {N - \frac{{f({{\bf{x}}_1})f({{\bf{x}}_2})}}{N}} \right.\\
&\left. { - \sqrt {{N^2} + \frac{{{f^2}({{\bf{x}}_1}){f^2}({{\bf{x}}_2})}}{{{N^2}}} - \left( {{f^2}({{\bf{x}}_1}) + {f^2}({{\bf{x}}_2})} \right)} } \right)\mathop  \ge \limits^{(5)} 0,
\end{split}
\end{equation}
where the inequality (4) is established since $\frac{{f({{\bf{x}}_1})}}{{\sqrt N }} - \frac{{f({{\bf{x}}_2})}}{{\sqrt N }} \ge 0$ and $\sqrt {N - \frac{{{f^2}({{\bf{x}}_1})}}{N}}  - \sqrt {N - \frac{{{f^2}({{\bf{x}}_2})}}{N}}  \le 0$, and then $\left( {\frac{{f({{\bf{x}}_1})}}{{\sqrt N }} - \frac{{f({{\bf{x}}_2})}}{{\sqrt N }}} \right)t + \left( {\sqrt {N - \frac{{{f^2}({{\bf{x}}_1})}}{N}}  - \sqrt {N - \frac{{{f^2}({{\bf{x}}_2})}}{N}} } \right)\sqrt {1 - {t^2}} $ is monotonically increasing w.r.t. $t \in \left[ {\frac{{f({{\bf{x}}_1})}}{N},1} \right]$, i.e., it achieves the minimum when $t = \frac{{f({{\bf{x}}_1})}}{N}$, and the inequality (5) is established since $N - \frac{{f({{\bf{x}}_1})f({{\bf{x}}_2})}}{N} = \sqrt {{N^2} + \frac{{{f^2}({{\bf{x}}_1}){f^2}({{\bf{x}}_2})}}{{{N^2}}} - 2f({{\bf{x}}_1})f({{\bf{x}}_2})} $ and further note that ${f^2}({{\bf{x}}_1}) + {f^2}({{\bf{x}}_2}) \ge 2f({{\bf{x}}_1})f({{\bf{x}}_2})$. Hence, $\Theta (t,{{\bf{x}}_2}) \le \Theta (t,{{\bf{x}}_1})$ when $t \in \left[ {\frac{{f({{\bf{x}}_1})}}{{\sqrt N }},1} \right]$.

Based on the above analysis, it is always favorable to increase the value of $f({\bf{x}})$ to maximize $\Theta (t,{\bf{x}})$. This thus completes the proof.
\end{proof}

  \begin{figure*}[b!]
 \vspace{-9pt}
  \hrulefill
\setcounter{mytempeqncnt}{\value{equation}}
\begin{equation} \nonumber
\begin{split}{}
\nabla {f_1}({{\bf{x}}^{(k)}}) = {\left[ {\frac{{\partial {f_1}({\bf{x}})}}{{\partial {x_1}}},...,\frac{{\partial {f_1}({\bf{x}})}}{{\partial {x_N}}}} \right]_{{\bf{x}} = {{\bf{x}}^{(k)}}}} = 2\kappa \left[ {\sum\nolimits_{i = 1}^N {\sin \left( {\kappa (x_i^{(k)} - x_1^{(k)})} \right)} ,...,\sum\nolimits_{i = 1}^N {\sin \left( {\kappa (x_i^{(k)} - x_N^{(k)})} \right)} } \right].
\end{split}
\end{equation}
\setcounter{equation}{\value{mytempeqncnt}}
\vspace{-12pt}
\end{figure*}

\subsection{Optimizing Antenna Positions}
Based on Proposition 2, to maximize $\Theta (t,{\bf{x}})$, we should first optimize antenna positions by considering the following problem\footnotemark \footnotetext{We select ${f^2}({\bf{x}})$ as the objective since it facilitates subsequent analysis. It is also possible to select the objective as $f({\bf{x}})$.}
 \begin{align}
&({\rm{P2}}):{\rm{  }}\mathop {\max }\limits_{{{\bf{x}}}} \ {f^2}({\bf{x}}) \buildrel \Delta \over = {\left| {\overline {\bf{h}} _{sr}^H({\bf{x}}){{\overline {\bf{h}} }_{sd}}({\bf{x}})} \right|^2}\tag{${\rm{8a}}$}\\
{\rm{              }}&\ {\rm{s.t.}} \quad (3{\rm{c}}), (3{\rm{d}}).\tag{${\rm{8b}}$}
 \end{align}

To solve (P2), we first expand its objective as
\begin{equation} \nonumber
\begin{split}{}
{f^2}({\bf{x}}) =& {\left| {\sum\nolimits_{i = 1}^N {{e^{j\frac{{2\pi }}{\lambda }{x_i}(\sin {\theta _{sd}} - \sin {\theta _{sr}})}}} } \right|^2}\\
 =& {\left( {\sum\nolimits_{i = 1}^N {\cos \left( {\kappa {x_i}} \right)} } \right)^2} + {\left( {\sum\nolimits_{i = 1}^N {\sin \left( {\kappa {x_i}} \right)} } \right)^2}\\
\end{split}
\end{equation}
\begin{equation} \nonumber
\begin{split}{}
 =& N + \underbrace {\sum\nolimits_{i = 1}^N {\sum\nolimits_{j = 1,j \ne i}^N {\cos \left( {\kappa ({x_i} - {x_j})} \right)} } }_{{f_1}({\bf{x}})} ,
\end{split}
\end{equation}
where $\kappa  = \frac{{2\pi }}{\lambda }(\sin {\theta _{sd}} - \sin {\theta _{sr}})$. Based on this expansion, (P2) can be equivalently formulated as
 \begin{align}
&({\rm{P2.1}}):{\rm{  }}\mathop {\max }\limits_{{{\bf{x}}}} \ {{f_1}({\bf{x}})} \tag{${\rm{9a}}$}\\
{\rm{              }}&\ {\rm{s.t.}} \quad \ \ (3{\rm{c}}), (3{\rm{d}}).\tag{${\rm{9b}}$}
 \end{align}

Note that (P2.1) is non-convex due to its complex objective. In the next, we exploit the successive convex approximation (SCA) technique to solve it. To proceed, we construct a convex surrogate function to locally approximate the objective by resorting to the second-order Taylor expansion \cite{Mawenyan_MIMO}. Specifically, given ${{\bf{x}}^{(k)}} = {[x_1^{(k)},...,x_N^{(k)}]^T}$ in the $k$th iteration, we can derive that
\begin{equation} \nonumber
\begin{split}{}
{f_1}({\bf{x}}) \ge& {f_1}({{\bf{x}}^{(k)}}) + \nabla {f_1}({{\bf{x}}^{(k)}})\left( {{\bf{x}} - {{\bf{x}}^{(k)}}} \right)\\
 &- \frac{\delta }{2}{\left( {{\bf{x}} - {{\bf{x}}^{(k)}}} \right)^T}\left( {{\bf{x}} - {{\bf{x}}^{(k)}}} \right) \buildrel \Delta \over = {f_2}({\bf{x}},{{\bf{x}}^{(k)}}),
\end{split}
\end{equation}
where $\nabla {f_1}({{\bf{x}}^{(k)}}) \in {{\mathbb{R}}^{1 \times N}}$ denotes the gradient of $\nabla {f_1}({\bf{x}})$ at ${{\bf{x}}^{(k)}}$, which can be easily derived as shown at the bottom of this page. In addition, the positive real number $\delta $ should be set to satisfy $\delta {{\bf{I}}_N}\underline  \succ  {\nabla ^2}{f_1}({\bf{x}})$ \cite{Mawenyan_MIMO}, where ${\nabla ^2}{f_1}({\bf{x}}) \in {{\mathbb{R}}^{N \times N}}$ is the Hessian matrix of ${{f_1}({\bf{x}})}$. Note that the element in the $i$th row and $i$th column of ${{\nabla ^2}{f_1}({\bf{x}})}$ is derived as ${\left[ {{\nabla ^2}{f_1}({\bf{x}})} \right]_{i,i}} =  - 2{\kappa ^2}\sum\nolimits_{j = 1,j \ne i}^N {\cos \left( {\kappa ({x_j} - {x_i})} \right)}$, $i = 1,...,N$. Similarly, the element in the $i$th row and $j$th column ($j \ne i$) of ${{\nabla ^2}{f_1}({\bf{x}})}$ is derived as ${\left[ {{\nabla ^2}{f_1}({\bf{x}})} \right]_{i,j}} = 2{\kappa ^2}\cos \left( {\kappa ({x_j} - {x_i})} \right)$. Then, since ${\left\| {{\nabla ^2}{f_1}({\bf{x}})} \right\|_2}{{\bf{I}}_N}\underline  \succ  {\nabla ^2}{f_1}({\bf{x}})$ and further
\begin{equation} \nonumber
\begin{split}{}
\left\| {{\nabla ^2}{f_1}({\bf{x}})} \right\|_2^2 \le& \sum\nolimits_{i = 1}^N {\sum\nolimits_{j = 1}^N {{{\left( {{{\left[ {{\nabla ^2}{f_1}({\bf{x}})} \right]}_{i,j}}} \right)}^2}} } \\
 \le &4{\kappa ^4}N{(N - 1)^2} + 4N(N - 1){\kappa ^4},
\end{split}
\end{equation}
we can select $\delta  = \sqrt {4{\kappa ^4}N{{(N - 1)}^2} + 4N(N - 1){\kappa ^4}} $ to strictly satisfy $\delta {{\bf{I}}_N}\underline  \succ  {\nabla ^2}{f_1}({\bf{x}})$.

 Via the above analysis, in the $k$th iteration, ${\bf{x}}$ can be optimized by solving the following problem
 \begin{align}
&({\rm{P2.2}}):{\rm{  }}\mathop {\max }\limits_{{{\bf{x}}}} \ {f_2}({\bf{x}},{{\bf{x}}^{(k)}}) \tag{${\rm{10a}}$}\\
{\rm{              }}&\ {\rm{s.t.}} \quad (3{\rm{c}}), (3{\rm{d}}),\tag{${\rm{10b}}$}
 \end{align}
which is convex and thus can be efficiently handled using the CVX tool.

\textbf{Complexity Analysis:} Denote the number of iterations for obtaining the stationary solution of (P2.2) as ${I_{\bf{x}}}$. In addition, the complexity of solving (P2.2) in each iteration is about ${\cal O}({N^{3.5}})$. Hence, the total complexity of solving (P2.2) is about ${\cal O}({I_{\bf{x}}}{N^{3.5}})$.

\textbf{Remark 1:} Note that when ${f^2}({\bf{x}})$ achieves its maximum, the correlation between $\rm{S}$-${\rm{U}}_1$ and $\rm{S}$-${\rm{U}}_2$ channels, which is defined as $\frac{1}{N}\left| {\overline {\bf{h}} _{s{u_1}}^H({\bf{x}}){{\overline {\bf{h}} }_{s{u_2}}}({\bf{x}})} \right|$, reaches the highest level. In other words, antenna positions at $\rm{S}$ should be first optimized based on the rule of maximizing the correlation between the above two channels. This finding, although novel, can be understood intuitively, since a higher correlation between $\rm{S}$-${\rm{U}}_1$ and $\rm{S}$-${\rm{U}}_2$ channels creates favorable prerequisites for effective beamforming designs to concurrently improve the received SNRs at ${\rm{U}}_1$ and ${\rm{U}}_2$.

\subsection{Optimizing Beamforming}
Denote the stationary solution of (P2.2) as ${f_{2,\max }}$. Hence, the stationary solution of ${f^2}({\bf{x}})$, denoted as ${f_{\max }^2}$, is expressed as $f_{\max }^2 = N + {f_{2,\max }}$. Now, substituting the known ${f_{\max }} = \sqrt {N + {f_{2,\max }}} $ and $f_{\max }^2$ into (6), the minimum received SNR will be only related to $t$ and is expressed as
\begin{equation}
\setcounter{equation}{11}
\begin{split}{}
\Theta (t) = \min \left( {{A_1}{t^2},{{\left( {{A_2}t + {A_3}\sqrt {1 - {t^2}} } \right)}^2}} \right),
\end{split}
\end{equation}
where ${A_1} = \frac{{{P_s}N}}{{d_{{s{u_1}}}^\tau {\sigma ^2}}}$, ${A_2} = \sqrt {\frac{{{P_s}}}{{d_{{s{u_2}}}^\tau {\sigma ^2}}}} \frac{{{f_{\max }}}}{{\sqrt N }}$ and ${A_3} = \sqrt {\frac{{{P_s}}}{{d_{{s{u_2}}}^\tau {\sigma ^2}}}\left( {N - \frac{{f_{\max }^2}}{N}} \right)} $. Based on (11), the remaining optimization problem can be formulated as
 \begin{align}
&({\rm{P3}}):{\rm{  }}\mathop {\max }\limits_{t} \ \Theta (t)\tag{${\rm{12a}}$}\\
{\rm{              }}&\ {\rm{s.t.}} \quad t \in \left[ {{f_{\max }}/N,1} \right].\tag{${\rm{12b}}$}
 \end{align}

 Problem (P3) can be easily solved by discussing three cases. Specifically, let ${y_1}(t) = {A_1}{t^2}$ and ${y_2}(t) = {\left( {{A_2}t + {A_3}\sqrt {1 - {t^2}} } \right)^2}$. Then, recall that ${y_2}(t)$ achieves its maximum when $t = \frac{{{A_2}}}{{\sqrt {A_2^2 + A_3^2} }} = \frac{{{f_{\max }}}}{N}$ and is monotonically decreasing w.r.t. $t \in \left[ {\frac{{{f_{\max }}}}{N},1} \right]$, while ${y_1}(t)$ is monotonically increasing w.r.t. $t \in \left[ {\frac{{{f_{\max }}}}{N},1} \right]$. Based on these we can easily determine that
 \begin{itemize}
 \item Case 1: If ${y_1}\left( {\frac{{{f_{\max }}}}{N}} \right) < {y_2}\left( {\frac{{{f_{\max }}}}{N}} \right)$ and ${y_1}(1) > {y_2}(1)$, the optimal $t$ must be the solution of the equality ${y_1}(t) = {y_2}(t)$, which implies that $t = \frac{{{A_3}}}{{\sqrt {{{\left( {{A_2} - \sqrt {{A_1}} } \right)}^2} + A_3^2} }}$.
\item Case 2: If ${y_2}\left( {\frac{{{f_{\max }}}}{N}} \right) < {y_1}\left( {\frac{{{f_{\max }}}}{N}} \right)$, the optimal $t$ is ${\frac{{{f_{\max }}}}{N}}$.
\item Case 3: If ${y_2}(1) > {y_1}(1)$, the optimal $t$ is 1.
 \end{itemize}

\textbf{Remark 2:} In our proposed scheme, ${\bf{x}}$ and ${\bf{w}}$ can be optimized separately, and thus the total complexity is about ${\cal O}({I_{\bf{x}}}{N^{3.5}})$. On the other hand, focusing on the original expression of $\min \left( {{\gamma _{{u_1}}}({\bf{w}},{\bf{x}}),{\gamma _{{u_2}}}({\bf{w}},{\bf{x}})} \right)$, if AO is exploited to solve (P1), given ${\bf{x}}$, we can derive the optimal and closed-form ${\bf{w}}$ using the method in Subsection B, by just replacing the objective of (P3) with $\min \left( {{\gamma _{{u_1}}}(t,{\bf{x}}),{\gamma _{{u_2}}}(t,{\bf{x}})} \right)$ shown in (5). While given ${\bf{w}}$, we can also find the convex surrogate functions for the terms ${\left| {\overline {\bf{h}} _{{s{u_1}}}^T({\bf{x}}){\bf{w}}} \right|^2}$ and ${\left| {\overline {\bf{h}} _{{s{u_2}}}^T({\bf{x}}){\bf{w}}} \right|^2}$ in (2), using the similar method in Subsection A, and then exploit the SCA technique to optimize ${\bf{x}}$, with the complexity of ${\cal O}({I_{{\bf{x'}}}}{N^{3.5}})$, where ${I_{{\bf{x'}}}}$ is the number of iterations. However, since ${\bf{w}}$ and ${\bf{x}}$ need to be optimized alternately, the total complexity of solving (P1) with AO would become ${\cal O}({I_{{\rm{out}}}}{I_{{\bf{x'}}}}{N^{3.5}})$, where ${I_{{\rm{out}}}}$ denotes the number of outer iterations. Note that except for the higher complexity, AO cannot not provide any insightful conclusions.

\vspace{-15pt}
\section{Simulation Results}
\vspace{-5pt}
\begin{figure*}
\vspace{-10pt}
\centering

\begin{minipage}{5.5cm}
\includegraphics[width=5.5cm]{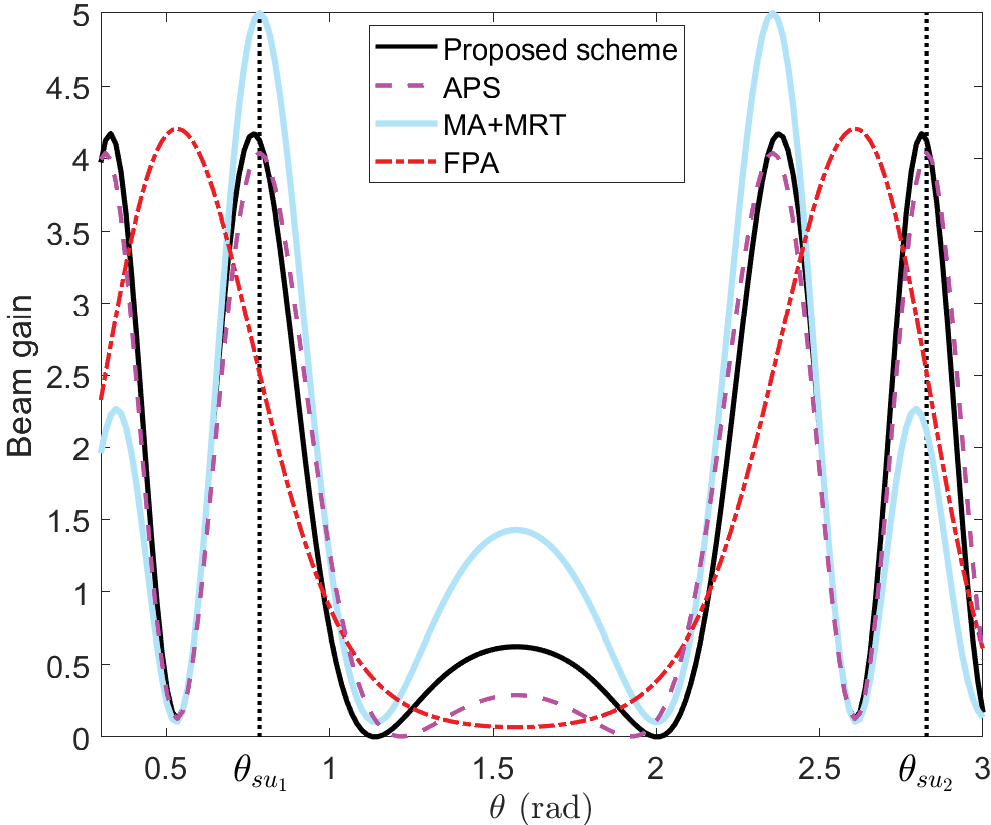}
\centering
\vspace{-8pt}
\subfigure{(a)}

\end{minipage}
\begin{minipage}{5.55cm}
\includegraphics[width=5.6cm]{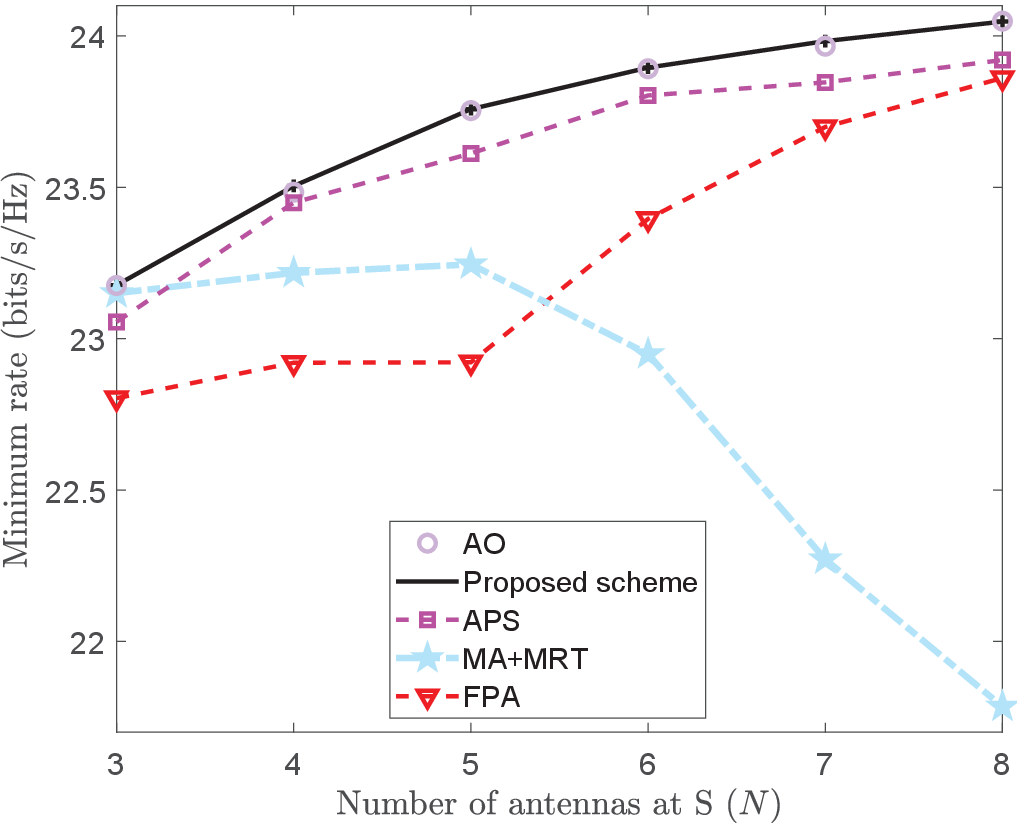}
\centering
\vspace{-8pt}
\subfigure{(b)}

\end{minipage}
\begin{minipage}{5.55cm}
\includegraphics[width=5.6cm]{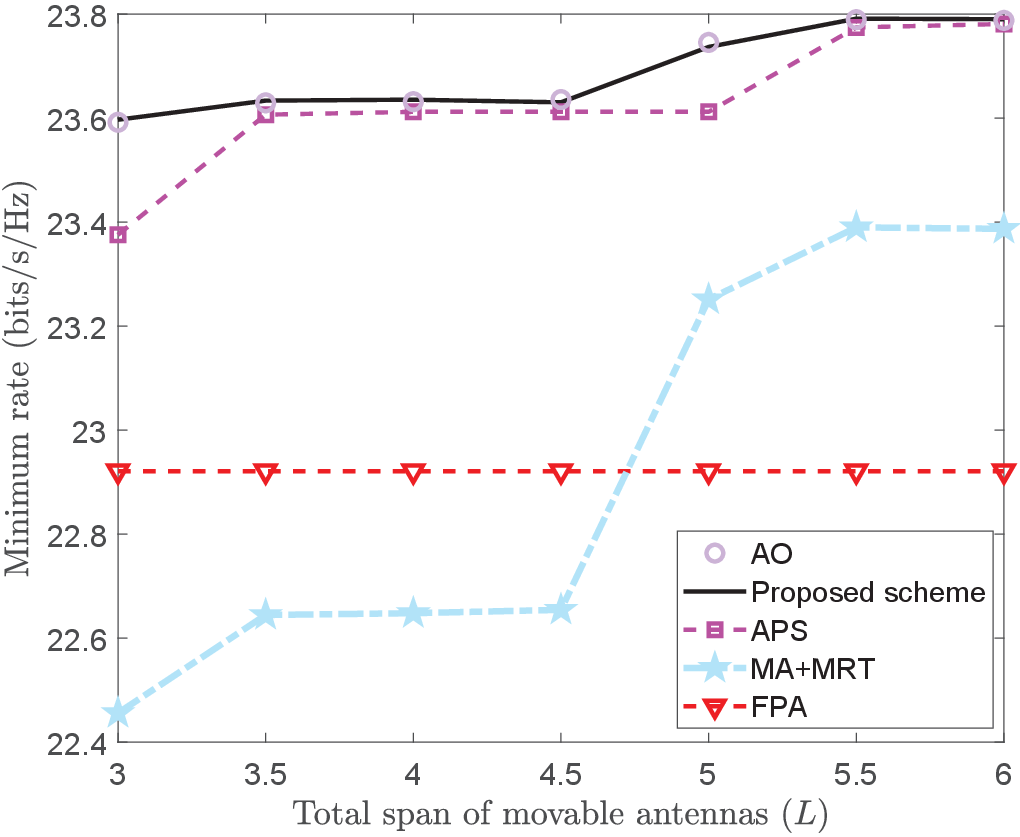}
\centering
\vspace{-8pt}
\subfigure{(c)}

\end{minipage}
\captionsetup{font=small}
\caption{(a) Beam gain of four schemes at different AoDs, $N = 5$ and $L = 4$; (b) Minimum rate versus the number of antennas at $\rm{S}$ ($N$), $L = 5$; (c) Minimum rate versus the total span of movable antennas at $\rm{S}$ ($L$), $N= 5$.}
\vspace{-20pt}
\end{figure*}

In this section, numerical results are presented to evaluate the effectiveness of the proposed scheme. Unless otherwise stated, the distance between $\rm{S}$ and ${\rm{U}}_1$ (${\rm{U}}_2$) is set as ${d_{s{u_1}}} = 100$ m (${d_{s{u_2}}} = 100$ m), the path loss exponent is set as $\tau  = 2$, the transmit power of $\rm{S}$ is ${P_s} = 25$ dBm, the noise power is ${\sigma ^2} =  - 80$ dBm, the carrier wavelength is normalized as $\lambda  = 1$, the minimum distance between two adjacent antennas at $\rm{S}$ is $d_{{\rm{min}}} = 0.5$, ${\theta _{s{u_1}}} = \pi /4$ and ${\theta _{s{u_2}}} = 9\pi /10$. For comprehensive comparisons, we consider four benchmarks:

1) AO: This scheme has been explained in Remark 2, and thus the details are omitted here;

2) Alternating position selection (APS): The total span of MAs is quantized into numerous discrete locations with equal-distance 0.5. The optimal position of each antenna is exhaustively searched from these locations to maximize the correlation between $\rm{S}$-${\rm{U}}_1$ and $\rm{S}$-${\rm{U}}_2$ channels. Then, $\rm{S}$ adopts the optimal beamforming described in Section IV.

3) MA+Maximum Ratio Transmission (MRT): $\rm{S}$ uses the proposed SCA technique to optimize antenna positions. Afterwards, $\rm{S}$ adopts the MRT-based transmit beamforming, i.e., ${\bf{w}} = {\bf{h}}_{s{u_1}}^*({\bf{x}})/\left\| {{\bf{h}}_{s{u_1}}^*({\bf{x}})} \right\|$, for information transmission.

4) FPA: The distance between arbitrary two adjacent antennas at $\rm{S}$ is fixed as 0.5. Under this setup, $\rm{S}$ adopts the optimal beamforming described in Section IV.

We first examine the beam gain of four schemes w.r.t. different AoD $\theta $ as shown in Fig. 2(a), where the beam gain at the angle $\theta $ with given ${\bf{w}}$ and ${\bf{x}}$ is denoted as ${\left| {\overline {\bf{h}} _{su}^T({\bf{x}}){\bf{w}}} \right|^2}$ with ${\overline {\bf{h}} _{su}}({\bf{x}}) = {\left[ {{e^{j\frac{{2\pi }}{\lambda }{x_1}\sin \theta }},...,{e^{j\frac{{2\pi }}{\lambda }{x_N}\sin \theta }}} \right]^T}$. From Fig. 2(a) we can observe that: i) for MA+MRT, even though antenna positions are carefully optimized to maximize the correlation between ${\rm{S}}$-${\rm{U}}_1$ and ${\rm{S}}$-${\rm{U}}_2$ channels, the beam gain at the direction where ${\rm{U}}_2$ is located will be very small, since the MRT-based beamforming only cares about the benefit of ${\rm{U}}_1$; ii) for FPA, even though the optimal beamforming is adopted to maximally balance the received SNRs at ${\rm{U}}_1$ and ${\rm{U}}_2$, the correlation between ${\rm{S}}$-${\rm{U}}_1$ and ${\rm{S}}$-${\rm{U}}_2$ channels cannot be proactively increased with FPAs, leading to the poor beam gain at the angles ${\theta _{s{u_1}}}$ and ${\theta _{s{u_2}}}$; iii) for APS, since antenna positions are optimized with the exhaustive search and concurrently the optimal beamforming is adopted, it achieves pretty good performance. However, since antennas can only be located in prescribed discrete points, the performance of APS is slightly inferior to our proposed scheme.

Fig. 2(b) further illustrates the minimum rate w.r.t. the number of antennas at $\rm{S}$ ($N$), from which we can observe that: i) as explained earlier, our proposed low-complexity scheme achieves the same performance compared to AO, verifying its effectiveness for practical implementations; ii) armed with the optimal beamforming design, as $N$ increases, there is a continuous performance improvement for the schemes of AO, proposed scheme, APS and FPA. While for MA+MRT, due to the improper beamforming design, the received SNR at ${\rm{U}}_2$ is proportional to ${\left| {\overline {\bf{h}} _{s{u_2}}^T({\bf{x}}){\bf{h}}_{s{u_1}}^*({\bf{x}})} \right|^2}/\left\| {{\bf{h}}_{s{u_1}}^*} \right\|_2^2 = {\left| {\overline {\bf{h}} _{s{u_2}}^H({\bf{x}}){{\bf{h}}_{s{u_1}}}({\bf{x}})} \right|^2}/{N^2}$, the value of which instead decreases as $N$ increases. Hence, the minimum rate achieved by MA+MRT becomes smaller when $N$ increases.

Fig. 2(c) shows the minimum rate w.r.t. the total span of MAs at $\rm{S}$ ($L$), from which we can observe that, for the schemes of AO, proposed scheme, APS and MA+MRT, as $L$ increases, antennas can be flexibly deployed in a larger space, i.e., the spatial DoF can be further explored to fully increase the correlation between ${\rm{S}}$-${\rm{U}}_1$ and ${\rm{S}}$-${\rm{U}}_2$ channels, the minimum rate of these schemes will become larger. Further, as $L$ continues to increase, the minumum rate of the above schemes will converge to a constant, indicating that it is not necessary to increase $L$ indefinitely and only a finite $L$ is enough to achieve the optimal performance. On the other hand, for FPA, since antenna positions are fixed regardless of what $L$ is, it achieves a constant rate w.r.t. $L$.
\vspace{-10pt}
\section{Conclusion}
This paper studies MAs-enabled two-user multicasting, and aims to maximize the minimum rate among two users by jointly optimizing the transmit beamforming and antenna positions at the source. Unlike most related works that adopt alternating optimization for iteratively handling the above two variables, we innovatively prove a new result, i.e., antenna positions can be first optimized based on the rule of maximizing the correlation between two multicasting channels; afterwards, the optimal beamforming can be computed. We believe that optimizing antenna positions from the perspective of varying channel correlations will open up a brand new path in MAs-enabled systems.

\normalem
\bibliographystyle{IEEEtran}
\bibliography{IEEEabrv,mybib}

\end{document}